\documentclass{./tlp} 
\usepackage[latin1]{inputenc} 
 
\usepackage{listings} 
\newenvironment{pcode}
    {\begin{lstlisting}[language=prolog,morekeywords=call,morekeywords=table]}
    {\end{lstlisting}}

\newenvironment{ccode}
    {\begin{lstlisting}[language=c,morekeywords=each]}
    {\end{lstlisting}}

\usepackage{graphicx} 
\usepackage{fourier} 
\usepackage[dvipsnames,usenames]{color} 

\newcounter{mnotei}  \newcommand{\mnote}[1]{%
{\scriptsize\textsf{\textcolor{blue}{$^{[\themnotei]}$}}}%
\marginpar{\scriptsize\textsf{\textcolor{red}{n.\themnotei: #1}}}%
\stepcounter{mnotei} } 
 
\renewcommand{\mnote}[1]{}

\newcommand{\redsect}{\vspace{-1em}}
\newcommand{\redfig}{\vspace{-1em}}

\begin{document} 
\bibliographystyle{acmtrans} 
 
\long\def\comment#1{}

\title[Memory-Scalable Answer-On-Demand Tabling]{Swapping
  Evaluation: A Memory-Scalable Solution for Answer-On-Demand Tabling}
 
\author[P. Chico de Guzm\'an et al.]{ 
  \begin{minipage}{0.3\linewidth}
\centering
Pablo Chico de Guzm\'an\\ 
\footnotesize
\emph{U.\ Polit\'ecnica de Madrid} \\ 
\emph{pchico@clip.dia.fi.upm.es} 
  \end{minipage}
\hfill
  \begin{minipage}{0.3\linewidth}
\centering
Manuel Carro\\ 
\footnotesize
\emph{U.\ Polit\'ecnica de Madrid} \\ 
\emph{mcarro@fi.upm.es}
  \end{minipage}
\hfill
\begin{minipage}{0.4\linewidth}
\centering
David S.\ Warren \\ 
\footnotesize
\emph{State University of New York at Stony Brook}\\ 
\emph{warren@cs.sunysb.edu}
\end{minipage}
} 
 

\pagerange{\pageref{firstpage}--\pageref{lastpage}} 
 
\maketitle 
 
\label{firstpage} 
 
\begin{abstract}  
  One of the differences among the various approaches to
  suspension-based tabled evaluation is the scheduling strategy.  The 
  two most popular strategies are \emph{local} and \emph{batched} 
  evaluation.   
  The former collects all the solutions to a tabled predicate before 
  making any one of them  available outside the tabled computation.
  The latter returns 
  answers one by one before computing them all, which in 
  principle is better if only one answer (or a subset of the answers) 
  is desired. 
  Batched evaluation is closer to SLD evaluation in that it 
  computes solutions lazily as they are demanded, but it may need 
  arbitrarily more memory than local evaluation, which is able to reclaim memory sooner.  
  Some programs which 
  in practice can be executed under the local strategy quickly run out 
  of memory under batched evaluation.  This has led to the general adoption of 
  local evaluation at the expense of the more depth-first batched strategy.
  In this paper we study the reasons for the high memory consumption of batched evaluation
  and propose a new scheduling strategy which we have termed
  \emph{swapping evaluation}.  Swapping evaluation also returns answers
  one by one before completing a tabled call, but its memory usage can
  be orders of magnitude less than batched evaluation.  An
  experimental implementation in the XSB system shows that swapping evaluation is
  a feasible memory-scalable strategy that need not compromise execution
  speed.
\end{abstract} 
 
\begin{keywords} 
  Logic Programming, Tabling, Implementation, On-Demand Answers, Performance. 
\vspace{-2em}
\end{keywords} 
 
\redsect
\section{Introduction.} 

Tabling~\cite{tamaki.iclp86-short,Warren92,chen96:tabled_evaluation}
is a strategy for executing logic programs that remembers subgoal
calls and their answers to respond to future calls.  This
strategy overcomes several of the limitations of the SLD resolution
strategy.  In particular, it guarantees termination for programs with
the bounded term size property and can improve efficiency in programs
which repeatedly perform some computation.  These characteristics help
make logic programs less dependent on the order of clauses and
goals in a clause, thereby bringing operational and declarative
semantics closer together.
Tabled evaluation has been successfully applied to deductive
databases~\cite{ramakrishnan93survey}, program
analysis~\cite{pracabsin,Dawson:pldi96-short}, semantic Web
reasoning~\cite{zou05:owl-tabling}, model
checking~\cite{ramakrishna97:model_checking_tabling-short}, etc.

One of the key decisions in the implementation of tabled evaluation is
when to return new answers to subsequent calls (called consumers),
i.e., the \emph{scheduling strategy}. \sloppy{Two main scheduling strategies
have been studied and applied so far: \emph{local} and
\emph{batched}~\cite{freire01:_beyon_depth_first}.}

Local scheduling computes \emph{all} the answers of a generator (the
first appearance of a call to a tabled predicate) before returning
them outside the generator subtree.  It is efficient in terms of time
and stack usage when all answers are needed. 
It is also efficient when an answer maximizing or minimizing some
metric is required, because usually \emph{all} the answers are
available when it comes to computing the extremal one.

Batched evaluation returns answers as soon as they are available.  It
is efficient for finding the first answer (or, in general, some but
not all answers) of a tabled predicate and for parallelizing tabled
evaluations: in parallel local evaluation, consumers outside the
generator subtree have to wait for the generator to complete, while in
batched evaluation these consumers could in principle run in parallel
and use answers stored in the table while the generator is still
computing and adding more answers.
Batched evaluation, however, may need arbitrarily more memory than
local evaluation: space can be reclaimed when subgoals have been
completely evaluated, and local scheduling completes goals earlier and
so can reclaim space earlier, while batched evaluation normally has
more subgoals in the process of being computed at any particular point
in time (and thus not completed).  Memory management is also
complicated by its usual stack-based nature, and batched evaluation
ends up with more unused memory trapped in the stacks.
We will analyze these factors and propose solutions to improve the
memory usage of batched evaluation while still keeping its good
behavior for first-answer scenarios.


The remainder of the paper is organized as follows:
Section~\ref{sec:motivation} gives a brief account of the advantages
that answer-on-demand tabling can bring.  Section~\ref{sec:overview}
gives an overview of tabled evaluation and the implementation of the
SLG-WAM in XSB.  Sections~\ref{sec:ASCC}
and~\ref{sec:external-consumers} explain why batched evaluation uses
more memory than local evaluation and propose how to improve it
through a new scheduling strategy, \emph{swapping
  evaluation}. Section~\ref{sec:performance} evaluates our solution
experimentally and shows that in some cases it uses orders of
magnitude less memory than batched evaluation without compromising
execution speed. As there are applications where swapping evaluation
is better than local evaluation and vice-versa,
Section~\ref{sec:simulate-local} suggests how to combine both
evaluations in the same engine. Finally,
Section~\ref{sec:impl-details} shows some implementation details. 


\redsect
\section{Motivation.}
\label{sec:motivation}

\mnote{Reviewer 1 comment} %
We can loosely divide many Prolog applications into two broad
\sloppy{categories: in-memory deductive} database applications where
all answers to a query are required, and artificial intelligence (AI)
search applications where it is required only to determine the
existence of a solution and perhaps provide an exemplar.  Tabled
Prolog has been effectively applied for the former type, such
as model checking and abstract interpretation, but not so effectively
for the latter, such as planning.  This may be traced back to the fact
that local scheduling is memory efficient for all-answer queries, but
batched scheduling, which is better for first-answer queries, shows
relatively poor memory utilization in the latter case.  XSB does
implement an optimization, called early completion, which is sometimes
able to avoid unnecessary search after a ground query has been found
to be true, but it is highly dependent on the syntactic form of the
rules and often allows unnecessary computation.  The swapping
evaluation strategy proposed in this paper is a depth-first search
strategy that has memory performance much closer to local scheduling.
In fact the order of its search is much closer to Prolog's order than
is batched scheduling, thus allowing Prolog programmers' intuitions on
efficient search orders to be brought to bear while programming in
tabled Prolog.  We believe swapping evaluation will make tabled Prolog
a much more powerful tool in tackling applications in AI areas
involving search.

\redsect\redsect
\section{An Overview of Tabled Evaluation.}
\label{sec:overview}

We assume familiarity with the WAM~\cite{hassan-wamtutorial} and
the general approach to implementing suspension-based tabled
evaluation, but for completeness we present an overview of 
the main tabling ideas in this section. We will focus on the
implementation approach of the 
SLG-WAM~\cite{sagonas98:xsb-abstract-machine} 
as it appears in XSB~\cite{xsb}, the platform on which we have developed
our prototype implementation.

Suspension-based tabling systems have four new operations beyond
normal SLD execution: \emph{tabled subgoal call}, \emph{new answer},
\emph{answer return}, and \emph{completion}.  \emph{Tabled subgoal
  call} checks if a call is a \emph{generator} (the first call to the subgoal) or a \emph{consumer} (a subsequent call).
If it is a generator, execution continues resolving against the
subgoal call clauses.  If it is a consumer, the \emph{answer return}
operation is executed and the consumer draws answers from an external
table, where the generator inserts each answer it finds using the
\emph{ new answer} operation.  

When no more answers are available for a consumer, its execution
suspends and freezes the stacks.  This is necessary since the
generator may generate new answers in the future, in which case the
consumer must be resumed to process that new answer.  Suspension is
performed by setting \emph{freeze registers} to point to then-current
stack tops.  No memory older than what the freeze registers point to
will be reclaimed on backtracking, since it may be needed to resume
the consumer.  Finally, when a generator has found all its answers, it
executes the completion operation, where relevant memory structures
are reclaimed and the freeze registers are reset to their original
values at the time of the generator call.  Tables are not reclaimed on
backtracking, and therefore do not need to be protected.

The completion operation is complex because a number of generators may
be mutually dependent, thus forming a Strongly Connected Component
(SCC~\cite{tarjan72}) in the graph of subgoal dependencies.  As new
answers for any generator can result in the production of new answers
for any other generator of the SCC, we can only complete all
generators in an SCC at once, when a fixpoint has been reached.  The
SCC is represented by the \emph{leader} node: the youngest generator
node which does not depend on older generators.  A leader node defines
the next completion point.

XSB implements a \emph{completion optimization} which obtains answers
directly from the table when a consumer returns answers from a
completed tabled subgoal.  With this optimization answers are not
necessarily returned in the original order. However, when answers are returned
from an incomplete tabled subgoal, they are returned in the original
order using an ordered list of answers.

Another important operation in the SLG\_WAM is consumer \emph{environment switching}.
When new answers are available for a suspended consumer, that consumer
is resumed to continue its suspended execution.  This is done by locating
the first common ancestor of the current execution point and the consumer to
be resumed.  Bindings from the current execution point to that ancestor
are undone, and bindings from the common ancestor until the consumer
are reinstalled.  
 
\redsect
\section{Improving Memory Usage by Precise Completion Detection.}
\label{sec:ASCC} 

One of the reasons for the importance of the completion instruction is
that it allows reclaiming memory from the WAM stacks. Before a
generator begins doing clause resolution, it saves the then-current
values of the 
freeze registers. Their values generally
change during the execution of the generator in order to preserve the
execution state of new consumers. When the generator completes, the
freeze registers are reset to their original values and
the frozen space is reclaimed.

\redsect
\subsection{An Overview of ASCC Memory Behavior.} 
\label{sec:ASCC-SCC} 

\begin{figure}[t]
  \begin{minipage}[b]{0.30\linewidth}
    \begin{pcode}
?- a(X).

@\neck@ table a/1, b/1.

a(X) @\neck@ b(Y), a(X). 
a(X) @\neck@ ... 
 
b(1). 
b(X) @\neck@ large, b(X).  @\end{pcode} 
  \vspace{2em}
  \caption{Tabled program.}
    \label{fig:tabled-program}
\end{minipage}
\hfill
\begin{minipage}[b]{0.66\linewidth}
  \centering\includegraphics[width=1.0\linewidth]{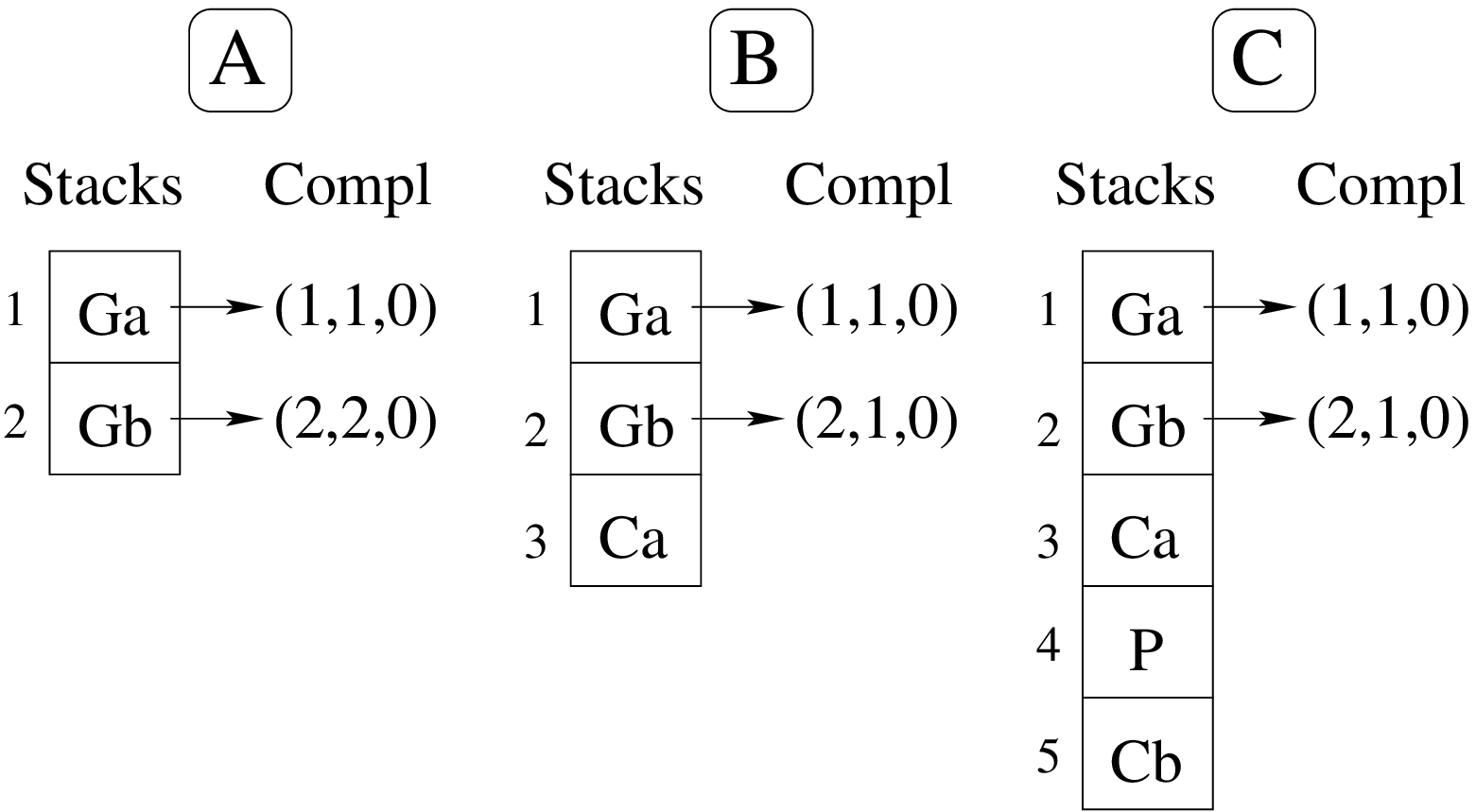}
  \caption{ASCC memory behavior.  ``P'' is a large SLD execution.}
  \label{fig:ASCC}
\end{minipage}
\redfig
\end{figure}

\mnote{This para. is v. diff. to understand.}
Batched evaluation, as implemented in XSB,
uses an approximation to detect \sloppy{SCCs (termed
  ASCC~\cite{sagonas98:xsb-abstract-machine}) in which} the completion
of some generators is postponed to ensure that memory needed by later
executions is not incorrectly reclaimed.  However, this results in
keeping some space frozen that could be reclaimed.  Consider the
example code in Figure~\ref{fig:tabled-program}.
Figure~\ref{fig:ASCC}(A) shows the stack of the generators, in which
\lstinline{Gb} is under \lstinline{Ga} at the moment of the \lstinline{b(Y)} call. The
triplets of the completion stack (on the right) are the original
identifier of each generator, its deepest dependence,\footnote{A
  generator can complete if its deepest dependence is not lower than
  its own identifier.} and the values of its freeze registers; note
that we are showing only the choicepoint stack, and therefore we need
just the freeze register for it.  Then, \lstinline{Gb} finds a solution
and a consumer (\lstinline{Ca}) of \lstinline{Ga} appears in the first
clause of \lstinline{a/1}.  Thus, \lstinline{Gb} cannot be completed before
\lstinline{Ga} when the ASCC is used to ensure that memory frozen by
\lstinline{Ca} is not incorrectly reclaimed (see
Figure~\ref{fig:ASCC}(B)).  \lstinline{Gb} continues its execution and a
consumer \lstinline{Cb} of \lstinline{Gb} appears, which freezes a lot of
memory.  That memory can be released upon completion of \lstinline{Gb},
but it is not released using the ASCC (see Figure~\ref{fig:ASCC}(C))
because the leader of \lstinline{Gb} is \lstinline{Ga}, and the completion
of \lstinline{Gb} is postponed.

\redsect
\subsection{A Solution: Imposing SCC Memory Behavior.}
\label{sec:ASCC-solution} 

The ASCC changes the structure of the SCC to overapproximate the
protection of frozen memory.  In the previous execution,   \lstinline{Gb} does not
depend on \lstinline{Ga}, because \lstinline{Ca}
lies outside the scope of 
\lstinline{Gb}.  Instead of changing the SCC structure, we propose to let
\lstinline{Gb} complete and release memory up to \lstinline{Ca} (the
youngest node belonging to a not-yet-completed SCC).  We do this by
changing the original values of the freeze registers that were stored when a
generator was created.  When a consumer appears, all the generators
that appear after that consumer's generator have their
original freeze register values updated to the current freeze register values.
Then, the memory frozen by the new consumer will not be reclaimed
if those generators complete.  

\begin{figure}
  \centering\includegraphics[width=0.66\linewidth]{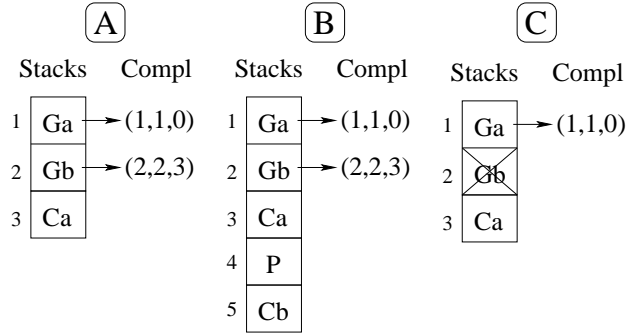}
  \caption{SCC memory behavior.}
  \label{fig:SCC}
\redfig
\end{figure}

Consider the previous execution using the new approach.  When
\lstinline{Ca} appears, the stacks are frozen until cell number 3
(included).  The generators which appeared after \lstinline{Ga} (the
generator of the new consumer) update their freeze register values to
be 3, but the SCC structure is not changed (see
Figure~\ref{fig:SCC}(A)).  Then, \lstinline{Cb} freezes the stacks
(see Figure~\ref{fig:SCC}(B)) but since \lstinline{Gb} is a leader
node, it can complete and the value of the freeze registers is updated
to be 3. All the memory frozen by \lstinline{Cb} is reclaimed
(including ancillary memory used by \lstinline{Gb} such as its ordered
list of answers
and list of consumers) without reclaiming the memory frozen by
\lstinline{Ca} (see Figure~\ref{fig:SCC}(C)).  The memory used by
\lstinline{Gb} is kept unreclaimed\footnote{It could, however, be
  collected by garbage collection.  This kind of \emph{trapped memory}
  already appears in the SLG-WAM and, interestingly, in the marker
  model for independent and-parallel execution when backtracking
  happens on \emph{trapped goals}~\cite{Backtracking} }.

\redsect
\section{Swapping Evaluation: the General Idea.} 
\label{sec:external-consumers} 
 
Both in local and batched evaluation, whenever a new consumer whose
generator has not been completed appears, execution suspends after
consuming all the available answers and WAM stacks are frozen. Some of
these suspensions are inherent to tabled evaluation, but some others
are not, as we will see in the next section.

\redsect
\subsection{External Consumers: More Stack Freezing than Needed.} 
\label{sec:external-problem} 
 
\mnote{New (clearer) definition?}We define two kinds of consumers:
\emph{internal} and \emph{external}. A consumer is internal to an SCC
when it appears inside the execution subtree of that SCC, and it is
external otherwise. As a simple example, let us have a program with
clauses \{\lstinline{(:-table a/0), (a :- a), (a)}\} and the query
\lstinline{?- a, a.} The leftmost \lstinline{a} in the query is a
generator, the \lstinline{a} in the body of the first clause is an
internal consumer, and the rightmost \lstinline{a} in the query is an
external consumer.
 
Freezing the stacks associated with internal consumers is necessary
for suspension-based tabled evaluation because they avoid infinite
loops.  But external consumers (which don't appear in local
evaluation) freeze the stacks so they can resume when the generator
produces additional answers. Note that new external consumers can
freeze the stacks again in those branches generated by a external
consumer suspension.
This stack freezing happens out of the scope of the tabled evaluation,
leading to a sort of memory-consuming breadth-first search which may
require significant memory as the following example shows:

\hspace{4em}
\begin{minipage}[b]{0.46\linewidth}
  \begin{pcode}
?- t(X), p, t(Y), fail.                       
@\neck@ table t/1.
t(1). 
t(2). @\end{pcode} 
\end{minipage}
\hfill
\begin{minipage}[b]{0.46\linewidth}
  \begin{pcode}
p @\neck@ large1. 
p @\neck@ large2. 
... 
p @\neck@ largeN. @\end{pcode} 
\end{minipage}

\lstinline{t(1)} is found and then the first clause of
\lstinline{p/0} is executed. Then \lstinline{t(Y)}, which is the first
consumer, consumes \lstinline{t(1)} and execution suspends after
freezing the stacks because there are currently no more available
answers for the consumer to return.  Execution backtracks to the
second clause of \lstinline{p/0} but the memory used by the execution
of the first clause of \lstinline{p/0} has not been reclaimed. The
same behavior will happen with the second consumer of \lstinline{t(X)}
after the second clause of \lstinline{p/0} succeeds. At the end of the
program, \lstinline{N} large computations corresponding to each of the
clauses of \lstinline{p/0} are frozen, in a fashion similar to a
breadth-first search evaluation. In the next section we will see how
this behavior can be avoided.

\redsect
\subsection{Swapping Evaluation: External Consumers No Longer Suspend.}
\label{sec:external-solution} 

An external consumer suspends when it does not have more available
answers, and it waits for its generator to (eventually) produce more
answers. We propose a different approach. When an external consumer
needs more answers, it is transformed into a generator to produce
them. Symmetrically, its generator is transformed into a consumer,
because answers will be computed by the new generator. We termed this
new scheduling strategy \emph{swapping} evaluation because an external
consumer and its generator are swapped. The swapping operation can be
seen as a change in the backtracking order since we backtrack over the
generator before backtracking over the top of the stacks.

Consider the previous example using swapping evaluation. When the
consumer \lstinline{t(Y)} needs more answers, it is transformed into
a generator to find the second answer
\lstinline{t(2)}.  Note that it does not recompute the first
  solution; it instead continues where the generator left the computation.
Then, execution fails (due to the call to \lstinline{fail/0} in the
query) and the swapped generator completes. The first clause of
\lstinline{p/0} is backtracked over, but now the space it used can be
reclaimed. A new consumer appears after the second clause of
\lstinline{p/0} succeeds which consumes both answers (using the
completion optimization) before \lstinline{fail/0} is
reached. Execution backtracks over the remaining clauses of
\lstinline{p/0} until it fails. Finally, the original
generator consumes the second answer \lstinline{t(2)}. The rest of the
execution continues as expected, with the former generator, now
consumer, \lstinline{t(X)} consuming saved answers. The result is that
at most one clause of \lstinline{p/0} is kept in the stacks at a time,
as in depth-first evaluation.

This is the most basic example of swapping evaluation, but the
swapping operation is complex due to swapping control, precise
completion detection, and the reordering of the stacks to change the
backtracking order. Section~\ref{sec:impl-details} gives
implementation details of swapping evaluation.
 


This scheduling strategy was prefigured in~\cite{JET-sagonas-PPDP},
where the authors suggested it as a way to recover SLD-like execution
to support cuts.  As they show, local and batched
evaluation do not follow the order of SLD resolution even if there are
no internal consumers, but swapping evaluation has an interesting
property: \emph{``if SLD resolution finishes, swapping evaluation
  keeps the clause resolution order of SLD, but some (redundant)
  branches are pruned.''}

We independently rediscovered this strategy by analyzing where memory
usage in batched scheduling was excessive and deriving methods to
improve that.  Our original contribution is the design of the
algorithms and data structures, their efficient implementation, and
their performance evaluation in XSB.  Our performance analysis
supports the effectiveness of the solution.

\redsect
\section{Experimental Performance Evaluation.} 
\label{sec:performance} 
 

We have implemented the techniques proposed in this paper in the XSB
system~\cite{xsb}.  All of the timings and measurements have been made
with XSB Version 3.2, disabling the garbage collector: we wanted to
study the effects in time and memory consumption of the different
evaluation strategies, without additional ``agents'' which could add
additional noise.
We used \lstinline{gcc 4.1.1} to compile the systems and we executed
them on a machine with Ubuntu 8.04, kernel 2.6.25, and an 1.6GHz Intel
Core 2 Duo processor.  Execution times are shown in ms.\ and memory
usage in bytes.

The benchmarks, which we will briefly explain here, are available from
\texttt{http://clip.dia.fi.upm.es/\~{}pchico/tabling/}.
\lstinline{tcl}, \lstinline{tcr}, and \lstinline{tcn} are transitive
closures on a graph with left, right, and double recursions,
respectively.  \lstinline{sg} is the well-known \emph{same generation}
program.  \lstinline{numbers} takes a list of numbers and a target
number \lstinline{N}, and tries to find an arithmetical expression
that evaluates to \lstinline{N} using operations from a fixed set and
\emph{all} the given numbers.  It uses both guided and blind search in
a potentially huge and irregular space, ultimately driven by number
theory.
\lstinline{atr2} is a compiled version of a probabilistic parser of
Japanese by Shigeru Abe, Taisuke Sato and Neng-Fa Zhou (13000+ lines), and
\lstinline{pg}, \lstinline{disj}, \lstinline{kalah},
\lstinline{gabriel}, \lstinline{cs_o}, \lstinline{cs_r} and
\lstinline{peep} are program analyzers created by automatically
specializing the original programs w.r.t.\ the generic
analyzers~\cite{codish98:analysis_with_XSB}, and whose sizes range
from 600+ to 1500 lines.

\redsect
\subsection{First-Answer Queries}
\label{sec:first-answer}

For queries requiring only one solution (see
Section~\ref{sec:motivation}), swapping/batched evaluation can be
significantly faster and use less memory (both in the stacks and in
the call/answer table) than local evaluation.
Table~\ref{tab:first-answer-comparison} shows the results, in time and
memory, of several such programs, under local and swapping evaluation.
Batched evaluation is not shown because it behaves quite similarly to
swapping evaluation in these cases.
The first four benchmarks look for the first answer of the query
presented in the table.
\lstinline{atr2_ground} parses a (ground) Japanese sentence of twelve
tokens. \lstinline{atr2_1var} and \lstinline{atr2_2var} parse the same
sentence but with the last token and the last and first token,
respectively, being free variables, which naturally leads to an
increase of the size of the search space.
\lstinline{numbers} is an example of the kind of program that merely
looks for a witness for the existence of a solution.  In
\lstinline{numbers_X}, \lstinline{X} represents the size of the set of
numbers.

\begin{table}[t] 
  \centering 
  \begin{tabular}{|l|r|r|r|r|r|r|}  
    \cline{1-7}
     & \multicolumn{3}{c|}{Local} & \multicolumn{3}{c|}{Swapping}  
    \\\cline{1-7} 
    Query & 
    Time & \parbox{5em}{\smallskip Stack  Memory}
    & \parbox{5em}{\smallskip Table Memory}
    & Time & \parbox{5em}{\smallskip Stack Memory}
    & \parbox{5em}{\smallskip Table Memory} 
    \\\cline{1-7}\cline{1-7} 
    tcl(100,\_)  & 0 & 2,320   & 1,828     & 0 & 3,276   & 212    
    \\\cline{1-7}
    tcr(100,\_)  & 20 & 162,756 & 89,108    & 0 & 21,164  & 9,852    
    \\\cline{1-7}
    tcn(100,\_)  & 20 & 190,660 & 90,640    & 0 & 2,228   & 212    
    \\\cline{1-7}
    sg(1,\_)    & 392 & 147,420 & 191,128   & 0 & 2,228   & 212    
    \\\cline{1-7} 
    atr2\_ground & 36 & 405,512 & 386,696   & 36 & 258,732 & 382,273
    \\\cline{1-7}
    atr2\_1var   & 1,048 & 522,844 & 3,864,540  & 744 & 296,244 & 3,374,687
    \\\cline{1-7}
    atr2\_2var   & 2,060 & 622,380 & 19,299,868 & 756 & 338,640 & 4,015,368
    \\\cline{1-7}
    numbers\_4   & 20 & 3,916   & 81,884    & 0 & 5,412 & 2,496    
    \\\cline{1-7}
    numbers\_5   & 544 & 7,108   & 2,406,312  & 1 & 6,632 & 4,416    
    \\\cline{1-7}
    numbers\_6   & 22,865 & 202,676 & 99,177,188 & 2 & 7,956 & 8,620    
    \\\cline{1-7}
  \end{tabular} 
  \caption{Time and memory comparison of local and swapping
    evaluations for first-answer queries.} 
\label{tab:first-answer-comparison} 
\redfig
\end{table} 

These results give a strong motivation for using answer-on-demand
tabling, as the behavior in time and memory of these benchmarks is
significantly better under swapping/batched evaluation than under
local evaluation.
%
Notice, specially, the exponentially bad behavior of
\lstinline{numbers} when local evaluation is used, while
swapping/batched evaluation remains linear.




\redsect
\subsection{All-Solution Queries.}
\label{sec:swapping-win}


We have also analyzed a set of well-known programs which are queried
to generate all the solutions.  
Note that, following the classification in
Section~\ref{sec:motivation}, these benchmarks exemplify the worst
case, where local evaluation, naturally devised to generate all the
solutions to a query, performs in general better that swapping
evaluation.  Therefore they are not representative of an average
behavior. 

\begin{table}[t] 
  \centering 
  \begin{tabular}{|l|r|r|r|r|c|} 
    \cline{1-6} 
    Program   & Local  & Batched-ASCC & Batched-SCC & Swapping & $\frac{Swapping}{Local}$
    \\\cline{1-6}
    tcl       & 2,248   & 2,176    & 2,708      & 2,172   & 0.97
    \\\cline{1-6}             
    tcr       & 196,068 & 180,368  & 180,692    & 178,768 & 0.91
    \\\cline{1-6}             
    tcn       & 229,392 & 209,648  & 209,972    & 208,644 & 0.91 
    \\\cline{1-6}             
    sg       & 764,960 & 813,276  & 813,600    & 790,068 & 1.03 
    \\\cline{1-6}             
    atr2      & 478,112 & 476,736  & 452,572    & 475,592 & 0.99 
    \\\cline{1-6}
    \multicolumn{6}{c}{}
    \\[-2ex]\cline{1-6}             
    pg        & 18,140  & 133,736  & 104,064    & 74,660  & 4.11 
    \\\cline{1-6}             
    disj      & 7,096   & 32,900   & 33,312     & 11,124  & 1.57 
    \\\cline{1-6}             
    kalah     & 11,700  & 61,060   & 39,944     & 23,324  & 1.99 
    \\\cline{1-6}             
    gabriel   & 20,256  & 42,460   & 42,268     & 22,700  & 1.12 
    \\\cline{1-6}             
    cs\_o     & 8,424   & 31,172   & 31,268     & 10,596  & 1.26 
    \\\cline{1-6}             
    cs\_r     & 9,532   & 31,896   & 28,976     & 11,420  & 1.20 
    \\\cline{1-6}             
    peep      & 22,700  & 354,612  & 77,664     & 78,572  & 3.46 
    \\\cline{1-6}
    \multicolumn{6}{c}{}
    \\[-2ex]\cline{1-6}             
    pg\_deep      & 18,564  & 339,744  & 307,572    & 32,384 & 1.74  
    \\\cline{1-6} 
    disj\_deep    & 23,852  & -       & 63,253,432  & 45,920 & 1.93  
    \\\cline{1-6} 
    kalah\_deep   & 29,116  & -       & -         & 132,232  & 4.54
    \\\cline{1-6} 
    gabriel\_deep & 30,884  & -       & 333,494,384 & 69,444 & 2.25  
    \\\cline{1-6} 
    cs\_o\_deep   & 8,356   & 63,420   & 50,680     & 32,988 & 3.95  
    \\\cline{1-6} 
    cs\_r\_deep   & 21,424  & 12,961,540 & 4,075,708   & 78,772 & 3.68  
    \\\cline{1-6} 
    peep\_deep    & 28,396  & -       & 51,194,340  & 106,228 & 3.74   
    \\\cline{1-6}                                                              
  \end{tabular} 
  \caption{Memory comparison for all-solution queries.} 
\label{tab:memory-comparison} 

\bigskip

  \begin{tabular}{|l|r|r|r|r|c|}  
    \cline{1-6} 
    Program   & Local   & B-ASCC  & B-SCC   & Swapping & $\frac{Swapping}{Local}$
    \\\cline{1-6}\cline{1-6}                       
    tcl       & 37.35 & 35.36 & 35.46 & 35.49 & 0.95
    \\\cline{1-6}                                            
    tcr      & 55.91 & 56.49 & 57.53 & 57.55 & 1.03
    \\\cline{1-6}                                        
    tcn      & 67.25 & 68.04 & 68.77 & 68.59 & 1.02
    \\\cline{1-6}                                        
    sg      & 263.27 & 272.57 & 275.99 & 293.37 & 1.11
    \\\cline{1-6}                                        
    atr2      & 872.64 & 876.29 & 884.07 & 884.65 & 1.01
    \\\cline{1-6}                                        
    \multicolumn{6}{c}{}
    \\[-2ex]\cline{1-6}             
    pg        & 9.02  & 8.93  & 9.043  & 9.06 &  1.00
    \\\cline{1-6}                                        
    disj      & 10.66 & 10.46 & 10.73 & 10.88 & 1.02
    \\\cline{1-6}                                        
    kalah     & 12.05 & 11.91 & 12.06 & 12.31 & 1.02
    \\\cline{1-6}                                        
    gabriel   & 13.48 & 13.19 & 13.45 & 13.52 & 1.00
    \\\cline{1-6}                                        
    cs\_o     & 18.54 & 18.49 & 19.03 & 18.82 & 1.02
    \\\cline{1-6}                                        
    cs\_r     & 36.43 & 36.54 & 37.21 & 36.57 & 1.00
    \\\cline{1-6}                                        
    peep      & 38.34 & 38.07 & 39.01 & 38.60 & 1.01
    \\\cline{1-6}                                    
    \multicolumn{6}{c}{}
    \\[-2ex]\cline{1-6}             
    pg\_deep      & 10.13 & 10.07 & 10.97 & 9.88 & 0.98  
    \\\cline{1-6}                                    
    disj\_deep    & 165.58 &    -    & 574.72 & 171.07 & 1.03 
    \\\cline{1-6}                                    
    kalah\_deep   & 17,375.49 &    -    &    -    & 17,147.00 & 0.99
     \\\cline{1-6}                                   
    gabriel\_deep & 2,749.17 &    -    & 4,180.91 & 2,808.98 & 1.02
    \\\cline{1-6}                                    
    cs\_o\_deep   & 1.88  & 1.81  & 1.87  & 1.94 &  1.03
    \\\cline{1-6}                                    
    cs\_r\_deep   & 70.27 & 79.51 & 99.15 & 69.36 & 0.99
    \\\cline{1-6}                                                           
    peep\_deep    & 273.05 &    -    & 384.99 & 272.07 & 0.99
    \\\cline{1-6}   \end{tabular} 
  \caption{Time comparison for all-solution queries.} 
\label{tab:time-comparison} 
\redfig
\end{table}

Their memory and time behavior appear, respectively, in
Tables~\ref{tab:memory-comparison} and~\ref{tab:time-comparison}.  In
both we show data for local evaluation, original batched evaluation,
batched evaluation with precise completion, and swapping evaluation
(which uses precise completion), plus a normalized comparison between
swapping and local evaluation.  We include both versions of batched
evaluation to determine whether the differences come from a more
precise SCC at completion or from using swapping evaluation.

We divide the benchmarks into three classes, according to their
structure. Benchmarks from \lstinline{tcl} to \lstinline{atr2} are
highly tabling intensive. They do not show big differences when using
the different tabling evaluations because they generate few SCCs. We
might in any case conclude that there is a slight
overhead due to the precise completion or/and the more involved
swapping control.
On the other hand, they in general favor swapping evaluation
memory-wise.

Benchmarks \lstinline{pg} to \lstinline{peep} call all the predicates
in the analyzers with free variables as arguments.  Every predicate is
called from a different clause and after the call finishes failure is
forced to generate all the solutions and to backtrack to the next
clause.  Solutions are kept in the table space and can be reused
between calls, as benchmark predicates call each other internally.
Forcing failure simulates a sort of local scheduling, even if the
engine supports swapping or batched evaluation.  For this reason,
local performs always better than swapping (but within reasonable
limits), and batched performs worse than swapping, but not with a huge
difference.  Precise completion brings advantages in some benchmarks.



The queries in the previous paragraph do not represent a common case
where there are few simultaneous dependencies between producers and
consumers.
Therefore, we have used a new category of queries where the program
code is the same as in the previous group, but queries to the
predicates in the analyzers are arranged in a conjunction, resulting
in a much more complex set of interdependences (again, due to
predicates internally calling each other). 
Generating all the solutions is, as before, forced by a \lstinline{fail/0}
call at the end of the conjunction.

In this category, swapping evaluation performs somewhat worse than
local evaluation both in memory and (with some exception) in time
behavior,\footnote{This is because swapping evaluation imposes some
  swapping control, some new data structure management, and mainly an
  execution stack reordering which leads to a non-negligible
  overhead. In any case, the differences are not very significant, and
  there are optimizations still available if execution speed proves to
  be a problem (see Section~\ref{sec:impl-details}).}  \mnote{hum?
  recheck footnote} due to the need to keep alive the environment
stacks of the generators in order to resume search for more solutions.
However, unlike batched evaluation (which is not even able to finish
some of the benchmarks), swapping evaluation maintains an acceptable
memory behavior.



As a conclusion, answer-on-demand tabling has been found to be
advantageous when only some answers are required.  However, batched
evaluation (the classical answer-on-demand strategy) was found to have
a very bad memory behavior in cases where complex dependencies appear
among tabled calls. This problem led to the use of local evaluation
for all applications (with the lack of efficiency in some cases), but
we think that our measurements indicate that swapping evaluation is a
viable alternative for answer-on-demand applications because it does
not have the bad-memory-behavior of batched evaluation.


\redsect
\subsection{Combining Local and Swapping Evaluation.}
\label{sec:simulate-local}

The previous section exhibits applications where swapping
evaluation performs much better than local evaluation (and vice-versa,
within reasonable limits).  While it could be possible to select the
adequate engine for every application, for simplicity, ease of
maintenance, and benefit of the final user, it would be nice to have
the tabling engine implement only one strategy.
We show that this is feasible by demonstrating how local evaluation can be
effectively emulated by swapping evaluation.  Assume that
\lstinline{t(X)} is a tabled predicate in a swapping evaluation
engine.  It is possible to generate automatically a wrapper which
evaluates \lstinline{t(X)} using local scheduling by:

\begin{enumerate}
\item Renaming the header of the clause(s) defining \lstinline{t/1} to be
  \lstinline{t_orig/1}.
\item Adding the following wrapper code:
\begin{pcode}
          t(X) @\neck@ call_is_consumer(t_orig(X)), !,  t_orig(X). 
          t(X) @\neck@ (t_orig(X), fail; t_orig(X)). @\end{pcode} 
\end{enumerate}

If the call to \lstinline{t(X)} is a consumer (determined using a
builtin available in the tabling engine), we consume from
\lstinline{t_orig(X)}, and we cut the second clause of
\lstinline{t(X)}.  Otherwise we force the generation of all the
answers for \lstinline{t_orig(X)} and then we consume
them.\footnote{Note that a similar transformation to make local
  evaluation behave as  batched or swapping
  does not seem to be possible.}

%
%
Experimentally, this simple simulation performs around 10\% worse than
local evaluation in memory and time when executing very intensive
tabling programs.  Note in this case the relatively costly swapping
operation is never invoked, since there are no external consumers.
Therefore the overhead comes from other sources (e.g., the
check/insert operations in the global table are executed three times
for every generator call: one for the \lstinline{call_is_consumer/1}
call, and two for the \lstinline{t_orig/1} calls of the second
clause), and 
a lower-level implementation should improve both memory and
time behavior.

In return, this transformation makes it possible to have, in the same
system and with the same engine, a predicate-level decision on whether
to evaluate under a local or a swapping policy, and use the
appropriate strategy in each case.  A similar consideration leads to
the combination of batched and local evaluation at the subgoal level
in~\cite{iclp/RochaSC05}.  However, in their work, batched evaluation,
with its disadvantages, is still used as the alternative to local
evaluation, and the way in which it is achieved is much more complex,
involving lower-level changes to the engine.



\redsect
\section{Swapping Evaluation Implementation Details.} 
\label{sec:impl-details} 

We now describe, at a somewhat high level, some implementation
details which provide an idea of the complexity inherent to the
implementation of swapping evaluation.

\redsect
\paragraph{\textbf{Generator Dependency Tree (GDT):} }
we use an XSB register named \lstinline{ptcp} (from \emph{parent
  tabled choicepoint}) which points to the nearest generator under
which we are executing, and which is stored in each consumer
choice point.  The \lstinline{ptcp} fields of the generator choice
points make up the GDT representing the creation order and
dependencies among generator calls.

\redsect
\paragraph{\textbf{Leader Detection:}}
we have added a new field to every generator choice point which keeps
track of the leader of each generator (\lstinline{NULL} if the
generator is a leader itself), which is used to accurately reconstruct
the SCC.  When a new consumer \lstinline{C} appears whose leader is
\lstinline{L$_\mathsf{C}$}, the generators in the current GDT branch
update their leader field to be \lstinline{L$_\mathsf{C}$}, until we
find a generator whose leader is already \lstinline{L$_\mathsf{C}$}.

\redsect
\paragraph{\textbf{External Consumers Detection:} }
a key to implement swapping
evaluation is determining whether a consumer is external or internal. To
do that, we have defined a new field in all the generator choice
points (the \lstinline{executing} field) which points to a free heap
variable. When the new answer operation is executed, that variable is
unified with some arbitrary value. Then, whenever a consumer appears,
if the \lstinline{executing} field of its leader generator points to a
free variable the consumer is internal and if that variable is
unified, the consumer is external. Note that that binding is undone if
we continue with the leader generator execution on backtracking, as we
need.


\redsect
\paragraph{\textbf{Creating Generators with Private Variables:} }
since generators can be swapped with external consumers, and the
execution segments of each generator can be moved to the top of the
stack, the \emph{tabled subgoal call} operation makes a private fresh
copy of the generator variables.  Then, all the bindings of the
generator call will be private to its execution, and those new
variables cannot be bound from outside the scope of the generator
execution.  Consequently, the execution subtree of the generator can
be moved to the top of the stacks in the same state as it was left.

An alternative possibility would be to untrail all bindings done
between the last answer of the generator and the external consumer
call in order to recover the original generator state each time an
external consumer (which was swapped by the generator) continues
making clause resolution to find new answers.  We have chosen to make
a private copy of generator variables because it does not require a
significant use of memory (between 2.5\% and 0.1\% in the benchmarks
we have executed) and it simplifies our implementation.  In terms of
speed, we cannot make strong conclusions because the \emph{untrailing}
alternative is not implemented, but we are quite confident that their
performance would be very similar.

Thus, in our implementation, each generator has two substitution
factors: one for the original generator call (to consume answers) and
another one for the answer bindings of private variables (to insert
them in the table). As a drawback, we lose the binding propagation of
batched evaluation which makes it faster than local evaluation in some
benchmarks.  On the other hand, swapping evaluation performs less trail
management (because external consumers do not switch their
environments) than batched evaluation, and, also, more consumers can
take advantage of the completion optimization because some external
consumers will find their table entry completed.

\redsect
\paragraph{\textbf{More Functionality in the \emph{New Answer} Operation:}}

Pointers to the tops of the trail and choice point stack are saved by
the \emph{new answer} operation when a generator finds a non-duplicate
answer. Two new fields of the generator choice points,
\lstinline{answer_cp} and \lstinline{answer_trreg}, remember those
values.  These pointers will be used to determine which parts of the
execution tree must be moved when the swapping operation is performed
to continue the execution from where it was left by the generator.

\redsect
\paragraph{\textbf{The Swapping Operation:}}
we term \lstinline{OldGen} the choice point of the generator and
\lstinline{NewGen} the choice point of the external consumer to be
swapped.  First, \lstinline{OldGen} is inserted into its corresponding
consumer list (the one belonging to the generator pointed to by the
\lstinline{ptcp} register of \lstinline{OldGen}) and \lstinline{NewGen}
  is erased from the consumer list it belongs to.  Then, the fields of
  the choice points are updated, such as the program counter, the
  substitution factor of the private copy of variables, the
  \lstinline{leader} of the new generator (\lstinline{NULL}), the last
  consumed answer of \lstinline{OldGen} (which is the last one
  found) and the \lstinline{executing} field of the new generator. The
  following code summarizes this operation:

\begin{ccode}
          PC(NewGen) = PC(OldGen);				
          PC(OldGen) = answer_return_inst;				
          PrivateSubstitutionFactor(NewGen) = PrivateSubstitutionFactor(OldGen);
          Leader(NewGen) = NULL;
          LastConsumeAnswer(OldGen) = LastAnswerFound(NewGen); 
          isExecuting(NewGen) = YES;@\end{ccode}
If \lstinline{OldGen} found an answer before leaving its execution
scope, we need to move to the top the segment of the choice point
stack which belongs to its clause resolution. In other words, we move
to the top all the choice points between the point where the last answer
of \lstinline{OldGen} was found (pointed to by \lstinline{answer_cp}) and
\lstinline{OldGen}. To do that, \lstinline{OldGen} will point to the
current top of the choice point stack and the choice point following 
\lstinline{answer_cp} will point to \lstinline{OldGen}.
This is implemented by scanning the choicepoints from
\lstinline{answer_cp} looking for a choice point which points to
\lstinline{OldGen}; that choice point is made to point to
\lstinline{NewGen}.\footnote{This traversal can be avoided by
  inserting a \emph{marker} choice point after \lstinline{OldGen} and
  updating its pointer to its previous choice point.}  The same
reordering is done with the trail\footnote{That means that trail cells
  are not any longer kept in relative order, and the way environment
  switching locates the first common ancestor of two consumers being
  switched has to be changed.  The new algorithm traverses the trail
  of each of the consumers alternatively, marking cells as they are
  traversed.  The first common ancestor is found when an
  already-marked cell appears.} and the local stack (indeed, the
program counter of the last local stack frame is also updated to point
to the continuation of \lstinline{NewGen}).

The next step is to reorder stacks from \lstinline{NewGen} to the
point where the last answer of the generator was found.  In this case,
reordering the local stack is not needed, and  the choice point
which points to \lstinline{answer_cp} is updated to point to
\lstinline{OldGen}.  The final result is that we have moved the
execution subtree of \lstinline{OldGen} to the top of the stack.

This stack reordering is also done for all consumers under the
execution of \lstinline{OldGen}, because they can belong to different
execution paths.\footnote{This step is not needed should the
  \emph{marker choicepoint} optimization be done, as all the consumers
  would be linked to that marker.}  Originally, we should traverse all
the generators in the same SCC \lstinline{OldGen} belongs to looking
for consumers which appear under the execution of \lstinline{OldGen}
by checking if \lstinline{OldGen} appears in their \lstinline{ptcp}
chain.  To make this checking more efficient, consumers are stored in the
list of consumers of the generator pointed to by their
\lstinline{ptcp} field (which is the nearest generator under they are
executing), instead of in the list of consumers of their
generator.\footnote{This change does not affect the rest of the
  tabling implementation and it should (heuristically) be more
  efficient than the original approach, because switching between
  consumers will be more likely to select those which are closer in
  the execution tree, thereby reducing the amount of work invested in
  untrail/redo operations.}

The final step is to reorder the completion stack and update the
freeze registers.  The portion of the SCC which \lstinline{OldGen}
belongs to and which is under the execution of \lstinline{OldGen}, has
to be moved to the top of the completion stack (because their
corresponding stacks have been moved to the top).  The freeze register
values of these generators are updated to protect the memory space of
\lstinline{NewGen} from backtracking.  This is because, as the
physical and logical order are different, after backtracking over a
choice point physically younger than \lstinline{NewGen}, the
associated memory to \lstinline{NewGen} would be wrongly reclaimed.
Indeed, the oldest generator among the generators younger than
\lstinline{OldGen}\footnote{The completion stacks gives us the age of
  the generators.} which does not belong to that SCC (called
\lstinline{G}) receives the original freeze register values of
\lstinline{OldGen}.  This is because when \lstinline{G} appears,
the freeze registers are protecting memory of the SCC which
\lstinline{OldGen} belongs to.  But that SCC was moved to the top
of the stacks, 
%
%
and the segment from the memory protected by the original freeze
register values of \lstinline{G} to the memory protected by the freeze
registers of \lstinline{OldGen} can be reclaimed when \lstinline{G}
completes.




\redsect
\paragraph{\textbf{A Sample Execution}}
\label{sec:sample-exec}

\begin{figure}[t]
  \centering
  \begin{minipage}[b]{0.29\linewidth}
    \begin{pcode}
?- a(X), inter, b(Y).

@\neck@ table a/1, b/1.

a(X) @\neck@ a(X).
a(X) @\neck@ code1, 
       b(X), 
       code2.

b(X) @\neck@ code3, 
       a(X), 
       code4.
b(X) @\neck@ code5, 
       X = 1. @\end{pcode}
  \caption{Non-trivial scenario.}
  \label{fig:realistic-scenario}
  \end{minipage}
     \hfill
  \begin{minipage}[b]{0.70\linewidth}
  \centering\includegraphics[width=1.05\linewidth]{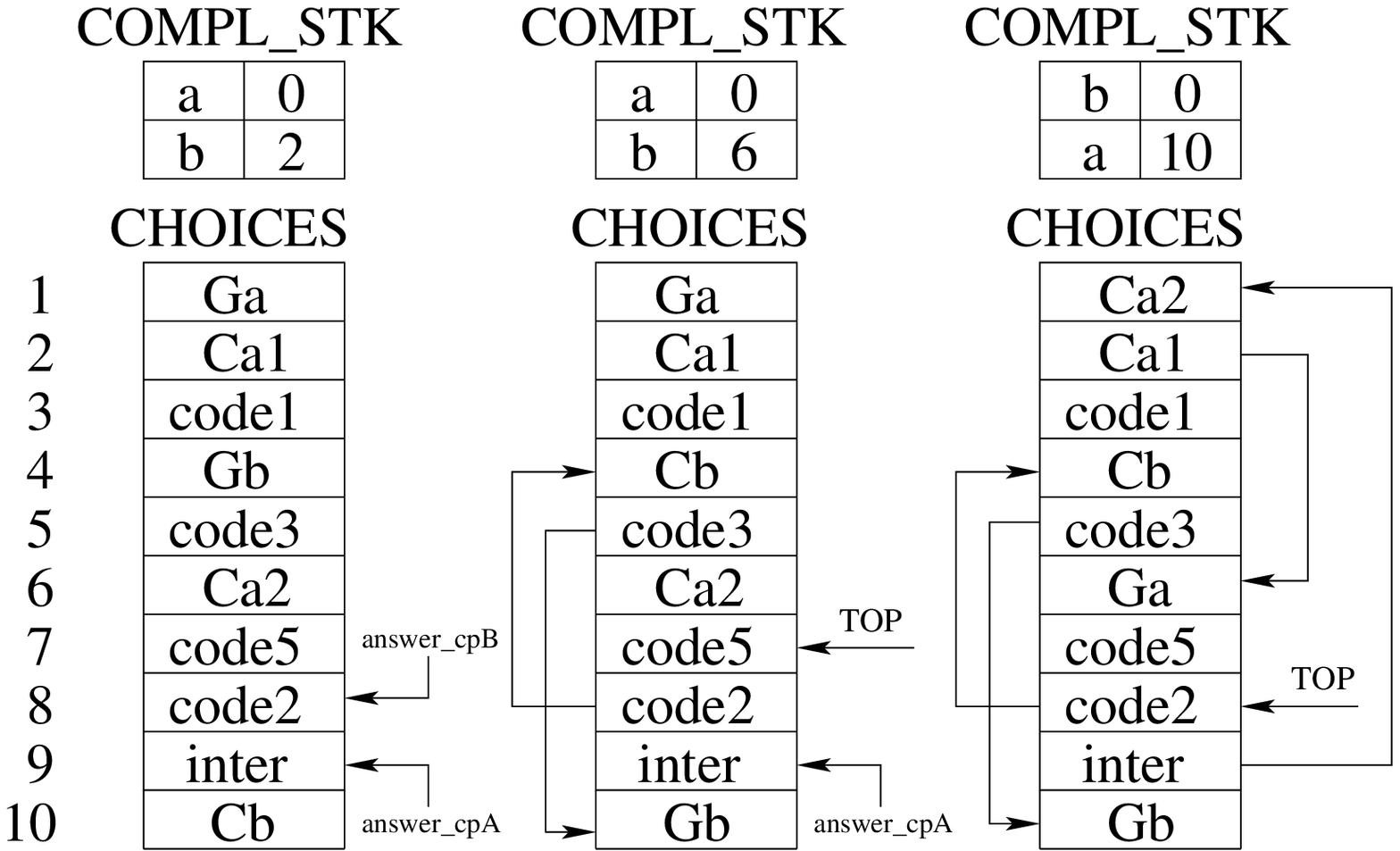}
  \caption{Choice point management.}
  \label{fig:choice-management}
  \end{minipage}
\redfig
\end{figure}

\begin{figure}[t]
  \centering\includegraphics[width=1.0\linewidth]{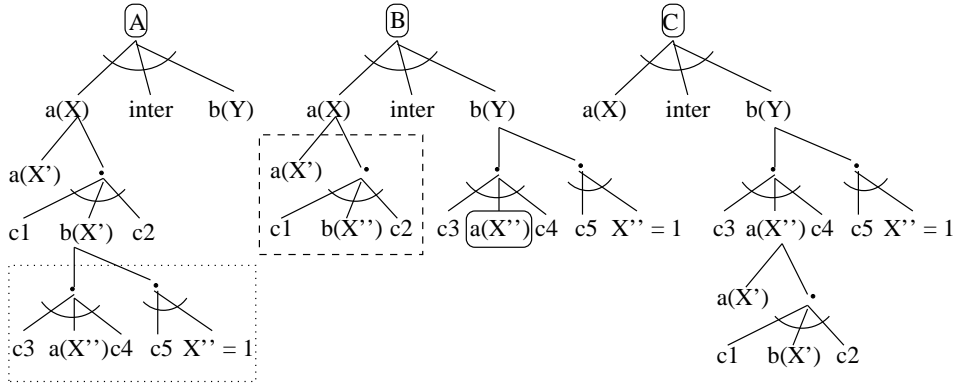}
\redsect
  \caption{And-Or tree execution.}
  \label{fig:and-or-tree}
\redfig
\end{figure}

We consider a non-trivial swapping evaluation scenario using the code
in Figure~\ref{fig:realistic-scenario}.  The first operation is the
tabled subgoal call of \lstinline{a(X)} where private variables
(marked with primes in Figure~\ref{fig:and-or-tree}) of the original
call are created to facilitate moving generator executions.  Then, an
internal consumer of \lstinline{Ga}, \lstinline{C$_{\mathsf{a1}}$},
appears and suspends. When the second generator, \lstinline{Gb},
creates its completion stack frame, the freeze register value is 2
(see Figure~\ref{fig:choice-management}(A)). A new internal consumer
of \lstinline{Ga}, \lstinline{C$_{\mathsf{a2}}$}, appears and
suspends, setting the freeze registers to 6. As swapping evaluation
needs a precise SCC (Section~\ref{sec:ASCC}), \lstinline{Gb} also
updates its original freeze register values. Later, \lstinline{Gb}
finds its first answer using its second clause. This answer is
propagated to the generator and \lstinline{Ga} finds its first answer
(Figure~\ref{fig:choice-management}(A) shows \lstinline{answer_cp} of
each generator).

After the execution of \lstinline{inter}, there is an external
consumer \lstinline{b(Y)} of \lstinline{Gb}.  When it consumes all the
available answers, a swapping operation is performed.  Using the limits
saved by the \emph{new answer} operation, the dotted rectangle in
Figure~\ref{fig:and-or-tree}(A) is moved under \lstinline{b(Y)} as
shown in Figure~\ref{fig:and-or-tree}(B). To do that, we update the
pointers to the previous choice point of \lstinline{code3} and
\lstinline{code2} and the current choice point is \lstinline{code5},
as shown in Figure~\ref{fig:choice-management}(B). A similar
reordering is done with trail cells and local stack frames. The new
generator \lstinline{Gb} saves the current values of the freeze
registers and the freeze registers are updated to be 10, to protect
the memory of \lstinline{Gb}.

The execution continues as expected until \lstinline{Gb} tries to
complete.  Note that the swapping operation has transformed
\lstinline{C$_{\mathsf{a2}}$}, which was an internal consumer, into an
external consumer.  Consequently, before checking for completion, all
the consumers under the execution of \lstinline{Gb} are checked in
case they have become external consumers.  We can easily access them
because, as explained before, consumers are saved in the consumer list
of the generator pointed to by their \lstinline{ptcp}.

Since \lstinline{C$_{\mathsf{a2}}$} is now an external consumer,
the swapping operation is performed to move the execution subtree in
the dashed rectangle in Figure~\ref{fig:and-or-tree}(B) under
\lstinline{C$_{\mathsf{a2}}$}, as shown in
Figure~\ref{fig:and-or-tree}(C). The reordered choice points are shown
in Figure~\ref{fig:choice-management}(C). As a consequence, generators
change their order in the completion stack. The new generator of
\lstinline{Ga} saves the value of the freeze registers and \lstinline{Gb}
takes their values from the previous values of \lstinline{Ga} to
reclaim all the memory upon completion of \lstinline{Gb}.


\redsect
\section{Conclusions.} 
 
We have presented swapping evaluation, a new strategy which retains
the advantages of batched evaluation such as first-answer efficiency
but which is memory-scalable without compromising execution speed.  We
have implemented swapping evaluation in XSB and experimentally tested
it in a series of benchmarks, with good memory and speed results.


The motivation behind this new evaluation strategy is to widen the
applicability of tabled Prolog from DB-like problems to other AI
applications, including e.g.\ search, where not all the solutions for a
given problem are required.


Finally, we believe that it would be advantageous to be able
to combine the advantages of local and swapping tabled evaluation in a
single system.  We have proposed a mechanism to easily simulate local
evaluation using swapping evaluation, which makes it possible to
define which evaluation to use at the predicate level.


\paragraph{\textbf{Acknowledgments:}} 
This work was funded in part by
IST-215483 grant {\em S-CUBE}, FET IST-231620 {\em HATS}, MICINN
project TIN-2008-05624 {\em DOVES}, and CM project P2009/TIC/1465
\emph{PROMETIDOS}.  Pablo Chico de Guzm\'an is also funded by an
Spanish FPU scholarship.
 
\bibliography{clip_bib/clip,clip_bib/general}

\label{lastpage} 
 
\end{document}